\documentclass[preprint,nofootinbib,aps,superscriptaddress]{revtex4}

\usepackage{graphicx}

\newcommand{\beq}{\begin{equation}}
\newcommand{\eeq}{\end{equation}}
\newcommand{\beqa}{\begin{eqnarray}}
\newcommand{\eeqa}{\end{eqnarray}}
\newcommand{\nn}{\nonumber}

\def\OMIT#1{{}}

\newcommand{\Bbar}{\,\overline{\!B}}
\def\B0bar{\Bbar{}^0}

\begin{document}

\preprint{\vbox{\hbox{MIT-CTP-3468} 
\hbox{CALT-68-2474} 
\hbox{hep-ph/0402076}
}}

\vspace*{1.25cm}

\title{\boldmath Stable Heavy Pentaquark States\vspace{0.2cm}}

\author{Iain W.\ Stewart\,}\email{iains@mit.edu}
\affiliation{Center for Theoretical Physics, Massachusetts Institute for
        Technology,\\[-4pt] Cambridge, MA 02139\vspace{4pt}  }

\author{Margaret E.\ Wessling\,}\email{wessling@theory.caltech.edu}
\affiliation{California Institute of Technology, Pasadena, CA 91125
        \\ $\phantom{}$ }

\author{Mark B.\ Wise\,\vspace{0.4cm}}\email{wise@theory.caltech.edu}
\affiliation{California Institute of Technology, Pasadena, CA 91125
        \\ $\phantom{}$  }

%\pacs{12.10.Dm, 12.10.Kt, 98.80.Cq}

\begin{abstract}
  
  If the $\Theta^+(1540)$ is interpreted as a bound state of a $\bar s$ quark
  and two $(ud)$ diquarks in a relative $P$-wave, then it is very likely that
  there exist pentaquark states with a heavy antiquark, $\bar b$ or $\bar c$,
  and two ``light'' diquarks in a relative S-wave which are stable against
  strong decays.  We make a mass estimate for exotic states of this type and
  discuss their weak decays. Isospin relations are constructed
  which test their flavor quantum numbers.

\end{abstract}
\maketitle

The $\Theta^+(1540)$ \cite{naknano} is a narrow state (with a width $\lesssim
15~{\rm MeV}$) that has the quantum numbers of $K^+n$. Since it has baryon
number one and strangeness minus one, it cannot be an ordinary $qqq$ baryon. A
promising quark model interpretation proposed by Jaffe and
Wilczek~\cite{Jaffe:2003sg} is that the $\Theta^+(1540)$ is a bound state of an
$\bar s$ quark with two $(ud)$ diquarks.\footnote{This state has also been
  interpreted in the chiral soliton model for baryons and large
  $N_c$~\cite{manohar}.}  The diquarks are in an isospin zero, spin zero, color
antitriplet configuration.  Since the diquark itself is a boson, there must be a
unit of orbital angular momentum between the two diquarks. Hence in this picture
the state has parity $P=+1$. Furthermore, part of the reason the
$\Theta^+(1540)$ is narrow can be explained by a small overlap with the
conventional $K^+n$ state~\cite{Jaffe:2003sg,carlson}.  A recent analysis of
$KN$ phase shifts suggests that $\Gamma_{\Theta^+} < 1.5\,{\rm
  MeV}$~\cite{Arndt}. Exotic cascades similar to the $\Theta^+$ but with
strangeness S=-2, can also be made (eg.  from two $(sd)$ diquarks and a $\bar
u$) and very recently there has been experimental evidence for these~\cite{alt}.
The pattern of strong decays for these exotics provides an interesting method
for distinguishing between different possibilities for the $J^P$ quantum
numbers~\cite{Jaffe:2003ci,Mehen}.

Jaffe and Wilczek~\cite{Jaffe:2003sg} also made a simple mass estimate
suggesting that states analogous to the $\Theta^+(1540)$, in which the $\bar s$
is replaced by a heavy antiquark, are bound. Denoting the states with flavor
structures $\bar c (ud)(ud)$ and $\bar b (ud)(ud)$ as $\Theta_{c}$ and
$\Theta_b$, their values $m_{\Theta_c}\simeq 2710\,{\rm MeV}$ and
$m_{\Theta_b}\simeq 6050\,{\rm MeV}$ are $100\,{\rm MeV}$ and $165\,{\rm MeV}$
below the strong decay thresholds to $pD^-$ and $n B^+$.  Given the
uncertainties, and that these estimates put the states fairly close to
threshold, this conclusion remains controversial. For instance in the model of
Karliner and Lipkin~\cite{karliner}, which has a triquark and diquark in a
P-wave, $m_{\Theta_c}\simeq 2985\,{\rm MeV}$ and $m_{\Theta_b}\simeq 6398\,{\rm
  MeV}$, so both are above the strong decay threshold.  Estimates based on the
constituent quark model also do not support their stability against strong
decays~\cite{Cheung}. On the other hand in a model with tensor
diquarks~\cite{Shuryak}, the masses come out close to those of Jaffe and
Wilczek~\cite{huang}.

There are in fact exotic pentaquark states containing a $\bar c$ or $\bar b$
quark that are more likely to be stable against strong decays in the diquark
picture than $\Theta_{b,c}$.  Diquark pairs of light $u,d$ and $s$ quarks come
in three types: $(ud)$, $(us)$, and $(ds)$, which form anti-triplets with
respect to SU(3) flavor and SU(3) color.  We will denote the interpolating field
for these diquarks by
\begin{equation} \label{notation}
  \phi^{a \alpha} = \epsilon^{abc}\epsilon^{\alpha \beta \gamma}
    q_{b\beta}q_{c\gamma} \,,
\end{equation}
where $\alpha, \beta$, and $\gamma$ are SU(3) color triplet indices and $a,b$,
and $c$ SU(3) flavor triplet indices (i.e., for $q_a$, $a=1$ corresponds to an
up quark, $a=2$ to a down quark and $a=3$ to a strange quark.)  In the
$\Theta^+(1540)$ the two diquarks are in the $\bar {\bf 6}$ representation of
$SU(3)$ flavor and the $\Theta^+(1540)$ is in a $\bar {\bf 10}$. With the
notation in Eq.~(\ref{notation}) its color and flavor quantum numbers are $\bar
s^\alpha \epsilon_{\alpha\beta \gamma}\phi^{3\beta}\phi^{3\gamma}$. For
$\Theta_{b,c}$ the states are in a $\bar {\bf 6}$ of SU(3) flavor. The heavy
pentaquark states we consider here have interpolating fields
\begin{equation}
  T_a =\epsilon_{abc}\,\epsilon_{\alpha \beta \gamma}\, \bar c^{\alpha}\, 
    \phi^{b \beta}\phi^{c \gamma}\,,\qquad
  R_a =\epsilon_{abc}\,\epsilon_{\alpha \beta \gamma}\, \bar b^{\alpha}\, 
    \phi^{b \beta}\phi^{c \gamma}\,.
\end{equation}
They are different from $\Theta_{c,b}$ since there is no P-wave between the
diquarks; hence the above states have parity $P=-1$ and are in the ${\bf 3}$
representation of $SU(3)$ flavor. Consider the case where the heavy quark is a
charm quark. Then as far as flavor quantum numbers are concerned these states
are $T_1=\bar c (ud)(su)$, $T_2=\bar c (ud)(sd)$ and $T_3=\bar c (su)(sd)$.
Because the two diquarks are distinct in each of these fields, the P-wave is not
required by Bose statistics.  To emphasize the strangeness and charge of the
states in this multiplet we will often use the notation $T_1=T_s^0$,
$T_2=T_s^-$, $T_3=T_{ss}^-$ for charm and $R_1=R_s^+$, $R_1=R_s^0$,
$R_3=R_{ss}^0$ for bottom. Here $\{T_s^0,T_s^-\}$ and $\{R_s^+,R_s^0\}$ form
isospin doublets, and $T_{ss}^-$ and $R_{ss}^0$ are isospin singlets.

The possibility of exotic pentaquarks $T_{1,2}$ was noted in
Refs.~\cite{Gignoux,Lipkin}, and the E791 collaboration has performed an
experimental search~\cite{Aitala:1997ja} in the mass range $2.75-2.91\,{\rm
  GeV}$. In Ref.~\cite{Lipkin} the states $T_{1,2}$ were called $P_{\bar c s}$.
In the context of the diquark model the exotic ${\bf 3}$ multiplet $T_a$ with
the diquarks in a relative S-wave was first discussed by Cheung \cite{Cheung},
but no mass estimate was given. Here we make a mass estimate for $T_a$ and $R_a$
in the diquark picture. We point out that the $R_a$ may be well below the strong
threshold, and the $T_a$ could be $\sim 200\,{\rm MeV}$ lighter than the E791
search window.  The exotic nature of these states can be determined through
measurement of weak decays, and we devise isospin and SU(3) relations which
could be used to further test their flavor quantum numbers if they are observed.

For our purposes, the most important feature of the $T_a$ and $R_a$ states is
that, within the diquark picture, they are more likely to be stable against
strong decays than $\Theta_{c,b}$, since they do not require the excitation
energy associated with a $P$-wave, ${\cal U}_{P\!-\!wave}$. To make a rough
estimate of their masses we write,
\begin{equation} \label{m1}
  m_{T_s}-m_{\Theta_c}= m_{R_s}-m_{\Theta_b}
  = \Delta_s -{\cal U}_{\rm{ P\!-\!wave}}
\end{equation}
where $T_s$ and $R_s$ denote the isodoublet states. Here $\Delta_s$ is the
change in mass from removing a $(ud)$ diquark and adding a $(us)$ or $(ds)$.
This takes into account that the $T_s$ ($R_s$) contain a strange quark unlike
the $\Theta_c$ ($\Theta_b$).  To estimate this term we use
\begin{eqnarray}
  \Delta_s \simeq m_{\Xi_c} - m_{\Lambda_c} = 184\,{\rm MeV} \,,
\end{eqnarray} 
and note that the $\Lambda$ and proton masses give a similar result
$\Delta_s\simeq m_{\Lambda}-m_p =177\,{\rm MeV}$.  A crude estimate of the
$P$-wave excitation energy is
\begin{equation}
\label{pwave}
  {\cal U}_{\rm{P\!-\!wave}} \simeq m_{\Lambda'_{c}}-m_{\Lambda_c} 
    = 310\,{\rm MeV}
\end{equation}
where $\Lambda'_{c}$ denotes the excitation of the $\Lambda_c$ with $(ud)$ in a
P-wave relative to $c$, and $m_{\Lambda'_{c}}=2594~{\rm MeV}$. This estimate is
supported by P-wave excitation energies for baryons built of light $(u,d,s)$
quarks, for example $m_{\Lambda(1405)}-m_\Lambda = 291\,{\rm MeV}$.
\OMIT{\footnote{For P-wave excitations of $(N,\Sigma,\Xi, \Lambda, \Delta)$ one
    finds ${\cal U}_{\rm{P\!-\!wave}}\simeq 300-500\,{MeV}$.}} We assign a
sizeable uncertainty to the estimate in Eq.~(\ref{pwave}) since it is easily
possible that it overestimates (or underestimates) the P-wave excitation energy
for $L=1$ between two diquarks. It also neglects possible Pauli blocking effects
between identical quarks in different diquarks.

Using Eq.~(\ref{m1}) and the Jaffe-Wilczek estimate,
$  m_{\Theta_c}=m_{\Theta}+m_{\Lambda_c}-m_{\Lambda} \simeq 2709~{\rm MeV}$,
gives the $T_s$ mass estimate 
\begin{equation} \label{masc} 
   m_{T_s} 
 %   \simeq  m_{\Theta}+2m_{\Lambda_c}-m_{\Lambda'_{c}}-m_p 
    \simeq  m_{\Theta}+m_{\Lambda_c}-m_{\Lambda} +m_{\Xi_c}-m_{\Lambda_c'}
    =  2580~{\rm MeV} \,.
\end{equation}
For $T_s\to D_s\, p$, the sum of the $D_s$ and proton masses is $2910~{\rm
  MeV}$, i.e. Eq.~(\ref{masc}) puts the state $330\,{\rm MeV}$ below threshold
and $170\,{\rm MeV}$ below the E791~\cite{Aitala:1997ja} search region.  A
similar analysis in the case where the heavy quark is a bottom quark gives the
mass formula
\begin{equation} \label{masb} 
  m_{R_s} 
  % \simeq m_{\Theta}+m_{\Lambda_b}+m_{\Lambda_c}-m_{\Lambda'_{c}}-m_p
  \simeq  m_{\Theta}+m_{\Lambda_b}-m_{\Lambda} +m_{\Xi_c}-m_{\Lambda_c'}
   =5920~{\rm MeV}
\end{equation}
which is $390\,{\rm MeV}$ less than the sum of the $B_s$ and proton masses.
Thus, even with the large uncertainties in the mass estimate in
Eqs.~(\ref{masc},\ref{masb}), it seems quite likely that these states are stable
against strong decays.  The extra strange quark in $T_{ss}$ and $R_{ss}$ will
increase their mass by about $\Delta_s$ relative to $T_s$ and $R_s$
respectively, so that 
\begin{equation} \label{masbcss}
 m_{T_{ss}} \simeq m_{T_s}+\Delta_s = 2770\,{\rm MeV} \,,\qquad
 m_{R_{ss}} \simeq m_{R_s}+\Delta_s = 6100\,{\rm MeV}\,.
\end{equation}
The presence of the extra strange quark does not make $T_{ss}$ and $R_{ss}$
closer to threshold since their strong decays must involve two $s$ quarks in the
final state, eg. $R_{ss}\to B_s\,\Lambda$.

The states $T_a$ (and $R_a$) are truly exotic, being baryons with charm $=-1$
(beauty $=1$) and strangeness $=-1,-2$.  In contrast, the lighter S-wave analog
of $T_a$ with $\bar c\to \bar s$ does not have exotic quantum numbers; it mixes
with excited nucleon states via annihilation and is therefore hard to detect.
For the charm and beauty exotics promising weak processes for detection are
$\bar c\to \bar s d\bar u$, $\bar b \to \bar c c \bar s$, and $\bar b\to \bar c
u\bar d$.  Typical nonleptonic decay channels such as $T_s^0\to p \phi \pi^-$
and $T_s^-\to \Lambda K^+\pi^-\pi^-$~\cite{Lipkin} always break up the diquarks.
This may substantially decrease the corresponding partial widths, particularly
if the narrowness of the $\Theta^+(1540)$ is partly due to such an effect. For
bottom this penalty can be postponed by decays to charmed exotics which preserve
the diquark correlation, for example $\Theta_b\to\Theta_c\pi$~\cite{llsw3}. In
our case the analogs are $R_a\to T_a \pi^+$ and $R_a\to T_a D^+$. Among the
two-body $\bar b\to \bar c u\bar d$ decays these exotic-to-exotic channels are
also dynamically favored by factorization~\cite{factor}, which favors producing
an energetic $u\bar d$ meson, and suppresses decays to energetic mesons built out
of other flavor combinations.

Assuming the $T_a$ and $R_a$ states decay weakly there are several
promising discovery channels. For charm these include
\begin{eqnarray} \label{Tlist}
  && T_s^0 \to \Lambda K^0,\, p\pi^-,\, p\phi\pi^-,\, \Lambda K^+\pi^-,\, 
   K^0 K^- p ,\, 
  %\phi K^0\Lambda ,\, K^0 K^+ \Xi^- \,,
   \\
  && T_s^- \to K^0\pi^- \Lambda,\, p\pi^-\pi^-,\, p \phi \pi^-\pi^- ,\,
   \Lambda K^+ \pi^-\pi^- \,,
   %K^0 K^0 \Xi^- 
   \nn\\
  && T_{ss}^- \to \Lambda\pi^-,\, \Xi^- K^0,\, \phi\pi^-\Lambda ,\,
  K^+\pi^-\Xi^-,\, K^0K^-\Lambda ,\, K^-\pi^- p,\, 
 % \phi K^0 \Xi^- ,\, K^0 K^0\Omega^- \,, 
  \nn
\end{eqnarray}
for bottom with $\bar b\to \bar c c\bar s$
\begin{eqnarray} \label{Rlist1}
  && R_s^+ \to J/\Psi p ,\, \bar D^0 \Lambda_c ,\, D^- \Sigma_c^{++} ,\, 
  \pi^- \Delta^{++} ,\, J/\Psi \phi p, J/\Psi K^+\Lambda ,\, D_s^- D_s^+ p,\,
  D_s^- K^+ \Lambda_c ,\\
  && \phantom{R_s^+ \to} 
      D_s^- K^0 \Sigma_c^{++} ,\, \bar D^0 D_s^+ \Lambda,\, 
      \bar D^0 \phi \Lambda_c ,\,
  \nn\\
  && R_s^0 \to K^0\Lambda,\, D^-\Lambda_c,\, \bar D^0 \Sigma_c^0 ,\, 
   \pi^- p ,\,   J/\Psi K^0\Lambda,\, D_s^- K^+\Sigma_c^0 ,\, 
   D_s^- K^0\Lambda_c ,\, \bar D^0 \phi \Sigma_c^0 , \nn\\
  && \phantom{R_s^0 \to }
  D^- D_s^+ \Lambda ,\, D^-\phi \Lambda_c \,, \nn\\
  && R_{ss}^0 \to \phi \Lambda ,\, J/\Psi \Lambda ,\, D_s^- \Lambda_c ,\,
   K^- p ,\, J/\Psi \phi\Lambda ,\, J/\Psi K^+ \Xi^- ,\, D_s^- D_s^+ \Lambda 
   ,\, D_s^- \phi \Lambda_c ,\,\bar  D^0 D_s^+ \Xi^- \,. \nn
\end{eqnarray}
and with $\bar b\to \bar c u \bar d$
\begin{eqnarray} \label{Rlist2}
  && R_s^+ \to D_s^- \Delta^{++},\, D_s^- \pi^+ p ,\, \bar D^0 \bar K^0 p,\, 
  \bar D^0 \pi^+ \Lambda ,\, D^- \bar K^0 \Delta^{++},\,
  \\
  && R_s^0 \to D_s^- p ,\, \bar D^0 \Lambda,\, D_s^- \pi^+ \Delta^0 ,\, 
  \bar D^0 \bar K^0 \Delta^0,\, D^- \bar K^0 p,\, D^-\pi^+\Lambda,\,
  \nn\\
  && R_{ss}^0 \to \bar D^0 \Xi^0 ,\, D_s^- \bar K^0 p,\, 
   \bar D^0 \bar K^0 \Lambda,\, \bar D^0 \pi^+ \Xi^-
   \,. \nn
\end{eqnarray}
In general the quantum numbers of the final states in
Eqs.~(\ref{Tlist}-\ref{Rlist2}) are not sufficient to tell us that the initial
state was exotic. Weak decays of $\Lambda_{b,c}$ and $\Xi_{b,c}$ can mimic some
of these channels through Cabbibo suppressed or penguin
transitions.\footnote{And in one case even Cabbibo allowed $\Lambda_b$ decays
  which mimic the $R_{ss}^0$ decaying through $\bar b\to \bar c c\bar s$.}
Exceptions are the $T_s^-$ decays in Eq.~(\ref{Tlist}) and the $R_s^+$ decays
$(\bar b \to \bar c u\bar d)$ in Eq.~(\ref{Rlist2}), since for these two cases
the final quantum numbers $ddd$ and $\bar c s uuu$ are sufficiently unusual.  In
other cases kinematic information is necessary. For two cases the required
information is fairly minimal, $T_{ss}^-$ ($\bar c\to \bar s d\bar u$) only has
contamination from weak $b$-baryon decays and $R_s^+$ ($\bar b\to \bar c c\bar
s)$ only has contamination from weak $c$-baryon decays.

The dynamics of nonleptonic decays is complicated, and some channels
may be suppressed relative to others. Therefore it is desirable to search in as
many channels as possible. If we consider measurements of multiple weak decays
then there are isospin relations between the nonleptonic decays of the $R_s$
states. For the $\bar b \to\bar c c\bar s$ transition
\begin{eqnarray}
 && \Gamma(R_s^+\to J/\Psi K^+\Lambda) = \Gamma(R_s^0\to J/\Psi K^0 \Lambda) \,,
  \quad
 \Gamma(R_s^+\to \bar D^0 \phi\Lambda_c) =\Gamma(R_s^0\to D^- \phi \Lambda_c) \,,
   \nn\\
 &&\Gamma(R_s^+\to D_s^- K^+ \Lambda_c) = \Gamma(R_s^0\to D_s^- K^0
 \Lambda_c)\,,\quad
  \Gamma(R_s^+\to \bar D^0 \Lambda_c) =\Gamma(R_s^0\to D^-  \Lambda_c) \,,\nn\\
 && 2\Gamma(R_s^+\to \bar D^0 \Sigma_c^+) =  \Gamma(R_s^+\to D^-\Sigma_c^{++})
  = \Gamma(R_s^0\to \bar D^0\Sigma_c^0) = 2\Gamma(R_s^0\to D^-\Sigma_c^+)\,,
 \nn\\
 &&\Gamma(R_{ss}^0\to \bar D^0 \Xi_c^0) = \Gamma(R_{ss}^0\to D^- \Xi_c^+) \,, 
\end{eqnarray}
while for $\bar b \to \bar c u \bar d$
\begin{eqnarray}
 && \Gamma(R_{ss}\to D_s^- \Delta^{++} K^-) 
  = 3\Gamma(R_{ss}\to D_s^- \Delta^+ \bar K^0) \,,\nn\\
 && \Gamma(R_s^+\to \Delta^{++} D_s^+) 
  = 3\Gamma(R_s^0 \to \Delta^+ D_s^-) \,.
\end{eqnarray}
For $\bar c\to \bar s d\bar u$ transitions of $T_{ss}$ the isospin relations may
be harder to test, for example
\begin{eqnarray}
 &&\hspace{-1cm}
  \Gamma(T_{ss}^-\to \phi \pi^0 \Sigma^-)
     =\Gamma(T_{ss}^-\to \phi \pi^-\Sigma^0) \,,\qquad
   \Gamma(T_{ss}^-\to \phi \bar K^0 \Delta^-) 
    = \Gamma(T_{ss}^-\to \phi  K^-\Delta^0)\,, \nn \\
 &&
  2\Gamma(T_{ss}^-\to  \bar K^0 \pi^0 \Delta^-)
     =3 \Gamma(T_{ss}^-\to  \bar K^0 \pi^-\Delta^0) \,.
\end{eqnarray}

There are also isospin and $SU(3$) relations between semileptonic decays. For
$T_a$ states decaying with a single baryon in the final state the isospin
relation, $\Gamma(T_s^0 \rightarrow p e \bar \nu_e) =\Gamma(T_s^- \rightarrow n
e\bar \nu_e)$, relates the Cabibbo allowed $T_s^0$ and $T_s^-$ decays.  For
decays with a baryon and a meson in the final state we find the following
isospin relations,
\begin{eqnarray}
  2\Gamma(T_s^0 \rightarrow \pi^0 p e \bar \nu_e) 
  &=&\Gamma(T_s^0 \rightarrow \pi^+ n e \bar \nu_e)
  =2\Gamma(T_s^- \rightarrow \pi^0 n e \bar \nu_e)
  =\Gamma(T_s^- \rightarrow \pi^- p e \bar \nu_e),
  \\
  2\Gamma(T_s^0 \rightarrow K^+ \Sigma^0  e \bar \nu_e) 
  &=&\Gamma(T_s^0 \rightarrow K^0 \Sigma^+ e \bar \nu_e)
  =\Gamma(T_s^- \rightarrow K^+ \Sigma^- e \bar \nu_e)
  =2\Gamma(T_s^- \rightarrow K^0 \Sigma^0  e \bar \nu_e),
  \nn\\
 \Gamma(T_s^0 \rightarrow \eta p  e \bar \nu_e)
  &=&\Gamma(T_s^- \rightarrow \eta n  e \bar \nu_e),
  \qquad 
 \Gamma(T_s^0 \rightarrow K^+ \Lambda  e \bar \nu_e)
  =\Gamma(T_s^- \rightarrow K^0 \Lambda  e \bar \nu_e).\nn \hspace{1cm}
\end{eqnarray}
On the other hand by using SU(3) we can derive relations involving $T_{ss}^-$.
The weak Hamiltonian transforms as a ${\bf 3}$ and we find
\begin{equation}
  2\Gamma(T_s^0 \rightarrow p e \bar \nu_e)
    = 3\Gamma(T_{ss}^- \rightarrow \Lambda e \bar \nu_e)
\end{equation}
There is also one (independent) SU(3) relation between the semileptonic decays
to a meson and a baryon for the strangeness -1 states:
\begin{equation}
  2\Gamma(T_s^0 \rightarrow \eta p e \bar \nu_e)
   -2\Gamma(T_s^0 \rightarrow K^+ \Lambda e \bar \nu_e)
  = \Gamma(T_s^0 \rightarrow K^0 \Sigma^+ e \bar \nu_e)
   -\Gamma(T_s^0 \rightarrow \pi^+ n e \bar \nu_e).
\end{equation}
Similar results can be derived for semileptonic $R_a$ decays.

The combinations of quarks $\bar Q suud$ and $\bar Q sudd$ do not have to have
$I=1/2$. For example an $S$-wave (ud) quark pair in a spin one configuration can
be a color antitriplet if it is in an $I=1$ configuration. Combining this with
the other $u$ quark gives the possibility of an $I=3/2$ final state. However, it
appears likely that these states are heavier than the states we have been
considering. Phenomenological evidence for this comes from the fact that the
$\Sigma_c$ is heavier than the $\Lambda_c$ (and the $\Sigma$ is heavier than the
$\Lambda$). These other isospin states will decay to the ones we are considering
via emission of a photon and if the mass splitting is large enough by emission
of a pion.

It is also possible to construct S-wave pentaquark states in which a heavy
quark, say a charm, is part of one of the diquarks.  The interesting states of
this type are part of the ${\bf \overline {15}}$ representation of SU(3); they
are
\begin{eqnarray}
  F_2^{11} &=& F_s^- = \bar u (ds)(cd)  \,,\qquad
  F_1^{22} = F_s^{++} = \bar d (su)(cu) \,, \qquad 
  F_3^{11} = F_{ss}^- = \bar u (ds)(cs) \,,\nn\\ 
  F_3^{22} &=& F_{ss}^+ = \bar d (su)(cs) \,,\qquad
  F_1^{33} = F^{++} = \bar s (ud)(cu)   \,,\qquad
  F_2^{33} = F^+ = \bar s (ud)(cd) \,.
\end{eqnarray}
The other members of this multiplet are unstable because they contain $q \bar q$
pairs of the same flavor. Unfortunately we are not able to draw conclusions
about their masses from the observed $\Theta$ mass. If they are stable against
strong decay the charged two states, for example, could be detected via the mode
$F^{++} \rightarrow p \pi^+$.
%\begin{eqnarray}
 %\Gamma(F_s^- \to \pi^- \Xi^- e^+ \nu) &=& 
 %  \Gamma(F_s^{++} \to\pi^+\Xi^0 e^+\nu) \,, \\
 %\Gamma(F_s^- \to K^- \Sigma^- e^+ \nu) &=& 
  % \Gamma(F_s^{++} \to \bar K^0 \Sigma^+ e^+ \nu) \,,\nn\\
 %\Gamma(F_{ss}^- \to K^- \Xi^- e^+ \nu) &=& 
  % \Gamma(F_{ss}^+ \to \bar K^0 \Xi^0 e^+\nu) \,,\nn\\
% \Gamma(F^{++} \to\eta p e^+\nu) &=& 
 %  \Gamma(F^+ \to \eta n e^+ \nu) \,,\nn\\
 %\Gamma(F^{++} \to K^+\Lambda e^+\nu) &=& 
  % \Gamma(F^+ \to K^0 \Lambda e^+\nu) \,,\nn\\
 %2\Gamma(F^{++} \to \pi^0 p e^+\nu) &=& 
  % \Gamma(F^{++} \to \pi^+ n e^+ \nu) = \Gamma(F^+ \to \pi^- p e^+ \nu) 
  % = 2\Gamma(F^+ \to \pi^0 n e^+ \nu) \,,\nn\\
 %2\Gamma(F^{++} \to K^+ \Sigma^0 e^+ nu) &=& 
 % \Gamma(F^{++} \to K^0 \Sigma^+ e^+ \nu) = 
 % \Gamma(F^+ \to K^0 \Sigma^0 e^+ \nu) 
 % = 2\Gamma(F^+ \to K^+ \Sigma^- e^+ \nu) \,. \nn
%\end{eqnarray}

In this brief paper we have noted that within the diquark picture there are
pentaquark states with a heavy antiquark that are more likely to be stable
against strong and electromagnetic decay than the $\Theta_{b,c}$. Their masses
are lower because Bose statistics does not force the diquarks to be in a
relative $P$-wave. While a truly quantitative mass estimate is not possible it
appears likely that the states $R$ containing a $\bar b$ quark, and $T$
containing a $\bar c$, are stable against strong decays.  We have examined some
of their possible decays and discussed some relations that follow from isospin
and SU(3) symmetry. A crucial aspect of their detectability is their production
rate via fragmentation of the heavy quark to these exotics.  A crude estimate,
inspired by the fact that $\Lambda_b$ production via fragmentation is a factor
of $\sim 0.3$ less than $B$ production, is that every additional quark (or
antiquark) costs a $0.3$. This suggests the production rate for the $R_s$ (or
$T_s$) states may be $\sim 10^{-2}$ of the $B_s$ (or $D_s$) mesons in agreement
with earlier estimates for pentaquark production~\cite{produce}.

\acknowledgments 
I.W.S.\ was supported in part by the Department of Energy under
cooperative research agreement DF-FC02-94ER40818 and by a DOE Outstanding Junior
Investigator award.  M.E.W.\ and M.B.W.\ were supported in part by the
Department of Energy under Grant No.~DE-FG03-92-ER40701.

\end{document}